\documentstyle[prl,aps]{revtex}

\begin{document}
\draft

\twocolumn[\hsize\textwidth\columnwidth\hsize\csname@twocolumnfalse\endcsname

\title{
On Halting Process of Quantum Turing Machine
}

\author{Takayuki Miyadera\cite{MIYA} and 
Masanori Ohya\cite{OHYA}
}

\address{Department of Information Sciences,
Tokyo University of Science,  Noda City, Chiba 278-8510,
Japan}

\date{\today}

\maketitle

\begin{abstract}
We prove that there is no algorithm to tell whether 
an arbitrarily constructed 
Quantum Turing Machine has same time steps for different 
branches of computation. 
We, hence, can not avoid the notion of halting to be 
probabilistic in Quantum Turing Machine.
Our result suggests that halting scheme
of Quantum Turing Machine and quantum complexity theory
based upon the existing halting scheme 
sholud be reexamined. 
\end{abstract}
\pacs{PACS numbers: 03.67.Lx}
]
 In \cite{Myers} Myers pointed out that there exists a 
problem in case different branches of quantum computation on a Quantum
Turing Machine (QTM) take different numbers of steps to complete their
calculation.
That is, observation of halting qubit may spoil 
the computation 
since it selects a branch of computation and the quantum
interference is destroyed after the selection. 
Subsequently several papers \cite{Ozawa1,LP,Shi} 
on the halting process of QTM were published.
In 
\cite{Ozawa1} Ozawa proposed a possible solution by means of quantum
nondemolition measurement. 
He showed that 
if one considers a restricted class of QTMs such that
the halting qubit and data slots are not changed
 after
the branch falls into the halting state,
the probability to obtain each
outcome by a given time does not depend upon the
fact whether one employed the halting protocol or not.
However there still remains a problem.
Even if one employs the protocol proposed, 
the notion of halting is still probabilistic.
That is, a QTM with an input sometimes halts
and sometimes does not halt. 
If
one can not get rid of the possibility of
such a probabilistic halting, one can not 
tell anything certain for one experiment
since one can not say whether an event of 
halting or non-halting occured with probability 
one or just by accident, say with probability $10^{-40}$.
Bernstein and Vazirani \cite{BV} deals only with 
QTMs whose branches halt at a same time or none of them halt.
%
If one restrict QTMs to such a class,
one does not have the problem anymore.
We will show, in the present letter, the restriction is
not realistic.
That is, we address the following question:
{\it for an arbitrarily given QTM and
an input, can we tell whether the halting is probabilistic or not?}
 We prove that the answer is negative.
It means that there exists no necessary and sufficient condition
to prescribe non-probabilistic halting QTMs.
Thus our result suggests that the probabilistic halting 
cannot be avoided and appears naturally.
\par

A QTM \cite{BV} consists of a head, a processor
and an infinite two-way tape with 
data slots and other (working) slots.
Thus the total Hilbert space is 
spanned by a complete orthonormalized
system $\{|x \rangle \otimes |\xi \rangle \otimes |q_j \rangle\}$, where
$x$ is an infinite sequence of the alphabets $\{B,0,1\}$ ($B$ is called
as blank)
with the condition that 
the number of non-blank cells is finite and $\xi \in {\bf Z}$ represents 
the head position and $q_j \in \{q_0,q_1,\cdots,q_N,q_f\}$ 
is an internal state. Here $q_0$ denotes an initial state and
$q_f$ a halting state. A QTM is constructed 
by assigning complex probability amplitudes 
(components of a unitary matrix) which satisfy local rule condition.
That is, only transitions between locally different configurations 
 are allowed.
According to \cite{BV},
the components are assumed to be computable complex number, since 
otherwise we can not construct the QTM. 
In QTM, the halting scheme is 
slightly different from classical 
Turing machine due to the reduction of wave packet \cite{Deu}.
For each step, we observe whether the internal state is $q_f$ or not,
in other words,
we perform the meaurement of the operator
 $|q_f\rangle \langle q_f|$.
When the outcome is $1$, we measure the data slots in the tape and 
recognize the computation result. When the the outcome is $0$,
we do not observe anything and proceed the computation.  
All the ever known effective computation shemes \cite{Shor,Grover}
halt with probability one at a certatin time and never halt before then. 
However for an arbitrarily constructed QTM, the different branches
of computation can have different numbers of computation steps in general.
In such a case the halting process or the notion of halting itself 
has the problem as described above. One way to avoid such a difficulty is 
considering only special type of QTMs and inputs. 
\par
We call a pair of a QTM $Q$ and its input $x$ {\it non-probabilistic
halting} iff
one of the following conditions is satisfied.
\\
i) There exists a $t_0 \in \{1,2,\cdots \}$ such that 
at step $t_0$, $Q$ under $x$ halts with probability one and 
for $s< t_0$ it halts with probability zero.
\\
ii) For all the steps, $Q$ under $x$ halts with probability zero.
\\
A pair of $Q$ and $x$ is called {\it probabilistic halting} when 
it is not non-probabilistic.
We ask whether there exists an algorithm to 
judge whether a QTM and an input 
is probabilistic or not \cite{note}.
\par 
To answer the above question, we 
assuume the existence of such an algorithm, (classical) 
Turing Machine (TM) $T_0$.
Then we can show contradiction. 
TM $T_0$ works so as to read input $\langle Q,x \rangle$ 
where $Q$ is a QTM and $x \in \{ 0,1\}^* $ is an input 
and determine whether $Q$ under the input $x$ is
probabilistic halting or not.
For reversible TMs, $T_1$ and $T_2$, 
let us define a special type of QTM $Q(T_1,T_2)$
which has two branches running $T_1$ and $T_2$ 
without interference as follows.
The internal state of $Q(T_1,T_2)$ consists of 
a doubly indexed set $\{ (q_*,j),(q_0,j),(q_1,j),
\cdots,(q_N,j),(q_f,j),(q_{*f},j)\}$,
where $j=1,2$ and $N$ is a sufficiently large number.
That is, the Hilbert space of the internal states holds
tensor product structure, ${\bf C}^{N+4}\otimes {\bf C}^2$.
The internal state is initialized with $|q_*,1\rangle$ 
and a halting state is $|q_{*f},1 \rangle$. 
$Q(T_1,T_2)$ with an input $y$ (finite string) behaves as
follows:\\
1) change the internal state from initial state
$|q_*,1\rangle$ to $\frac{1}{\sqrt{2}}(|q_0,1 \rangle +|q_0,2 \rangle)$\\
2) for the branch with the second qubit of internal state $|1\rangle$,
execute the TM $T_1$ under the input $y$ 
and for the branch of $|2\rangle$, execute $T_2$ under the unput $y$.\\
3) If the internal state is $|q_f,j \rangle\ (j=1,2)$, change the 
internal state plus a fixed tape working cell  
into $|q_{*f},1 \rangle \otimes |j\rangle$. (i.e., To 
satisfy unitarity, the information
which branch was lived in is transferred to the tape cell.)
\par 
Denote the set of all the QTMs of above type as $S$, i.e.,
$S:=\{Q(T_1,T_2)|\ T_1,T_2 \mbox{ are reversible TMs}\}$.
Since $S$ is a subset of whole set of QTMs, TM $T_0$ could  
determine whether or not  $Q(T_1,T_2)$ under the input $x$ 
is probabilistic halting.
That is, we can determine that for any given reversible TM $T_1$ and 
$T_2$ their computing times for an input $x$
are the same or not. 
Thus we obtain a TM $T'_0$ which reads input 
$\langle T_1,T_2,x \rangle$ to  
compare their computing times, whose output is
"Yes" if their computing times are same and 
otherwise "No".
\par 
 By use of $T'_0$, we can construct the following TM $T_f$ with its 
 input $\langle T_1, x\rangle$ 
 where $T_1$ is a reversible TM and $x$ is its input.\\
 i) Read $T_1$ and $x$ \\
 ii) Construct a TM $T_2$ which never halts under any input \\
iii) input $\langle T_1,T_2,x\rangle$ to $T'_0$ \\
iv) Write the output of iii)
\par
We can see that if the outcome is "Yes" TM $T_1$ under the input $x$
does not halt and if the outcome is "No" TM $T_1$ under the input $x$
halts.
It contradicts the undecidability of 
halting problem \cite{Penrose} of classical TM. 
Thus our assertion was proved.
 \par
Here we proved that for an arbitrarily constructed QTM we can not 
say whether it is probabilistic halting or not. 
The result will suggest that to consider 
QTMs with diffierent computation steps for each branches is
necessary and the notion of halting in QTM should be reexamined
again. For instance it may play an important 
role to construct a quantum version of algorithmic complexity theory.
\par
T.M. thanks Masanao Ozawa, Fumihiko Yamaguchi and Satoshi Iriyama for 
helpful discussions.


\begin{references}
\bibitem{note}
Of course some special QTMs can be shown to be
probabilistic or non-probabilistic. 
However what we want to know is the existence of 
some universal algorithm independent of 
QTMs.
\end{references}
\end{document}